# The Mutual Information of University-Industry-Government Relations: An Indicator of the Triple Helix Dynamics

Loet Leydesdorff

Science & Technology Dynamics, University of Amsterdam

Amsterdam School of Communications Research (ASCoR)

Kloveniersburgwal 48, 1012 CX Amsterdam, The Netherlands

loet@leydesdorff.net ; http://www.leydesdorff.net/

## Abstract

University-industry-government relations provide a networked infrastructure for knowledge-based innovation systems. This infrastructure organizes the dynamic fluxes locally and the knowledge base remains emergent given these conditions. Whereas the relations between the institutions can be measured as variables, the interacting fluxes generate a probabilistic entropy. The mutual information among the three institutional dimensions provides us with an indicator of this entropy. When this indicator is negative, self-organization can be expected. The self-organizing dynamic may temporarily be stabilized in the overlay of communications among the carrying agencies. The various dynamics of Triple Helix relations at the global and national levels, in different databases, and in different regions of the world, are distinguished by applying this indicator to scientometric and webometric data.

# 1. Introduction

In 1953, Linus Pauling and Robert B. Corey proposed that DNA was made up of three chains, twisted around each other in ropelike helices (Pauling & Corey, 1953). A few months later, James Watson and Francis Crick proposed the double helix, which was then quickly accepted as the correct structure of DNA (Watson & Crick, 1953). This discovery led to a Nobel Prize (Watson, 1970).

Double helices can under circumstances stabilize in a coevolution, but triple helices may contain all kinds of chaotic behaviour (Poincaré, 1905). Triple Helix models continue to be useful in studying transition processes, for example, in crystallography and molecular biology. More recently, Richard Lewontin (2000) used the metaphor of a Triple Helix for modeling the relations between genes, organisms, and environments.

In a different context, Henry Etzkowitz and I introduced a Triple Helix model for the dynamics of university-industry-government relations (Etzkowitz & Leydesdorff, 1995). Our argument for using this neo-evolutionary model was that a knowledge-based regime of innovations can be expected to remain in transition. A Triple Helix can contain double helices as temporary stabilizations, but a system of three dynamics is meta-stabilized. Under specific conditions the next-order system of an overlay of communications can also be globalized and then exhibit self-organization. Globalization means in this context that the next-order (emerging) overlay gains priority in determining the dynamics of the underlying ones (on which it rests). Thus, a Triple Helix model may be sufficiently complex to encompass the different species of observable behaviour in the networks under study.

The advantages of using the Triple Helix model can be specified with reference to different research traditions. First, one is able to study specific configurations of university-industry-government relations as *instantiations* of the Triple Helix dynamics of a knowledge-based innovation system (Giddens, 1984; Leydesdorff & Etzkowitz, 1998). In this context of specification, the Triple Helix metaphor functions as a heuristics. The institutional configurations in knowledge-based systems can be considered as the outcome of three (functional) subdynamics of competitive systems: (a) the economic dynamic of wealth generation through exchange, (b) the knowledge-based dynamic of reconstruction and innovation over time, and (c) the political and managerial need and urge for normative control at the interfaces. The carriers of these three functions do no longer have to exhibit a one-to-one correspondence to industry, university, and government, respectively. The institutions can be expected to experiment with new formats in their mutual arrangements (Etzkowitz & Leydesdorff, 1997).

While the heuristic application of the Triple Helix metaphor can be made useful for the historical specification, the neo-evolutionary model of the two layers of functions and institutions operating upon each other opens a space of possible interactions. The evolutionary system has an option to reconstruct itself in the present with reference to the historical configurations that have occurred. The functional dimension can be provided with priority if a next-order system (e.g., a relevant selection environment) can be defined. Are the institutional arrangements still functional?

For example, participants who are entrained in co-evolutions of mutual shaping between two helices can be expected to 'lock-in' (David, 1985; Arthur, 1988). The internal perspectives of these participant-observers can be distinguished from the perspective of an external (that is, third) observer. The latter is able to evaluate. The switch to the external perspective enables the analyst to search for options emerging from interactions that cannot be perceived from within the co-evolution. The configuration under study can then be reconstructed on the basis of *knowledge*. Thus, a knowledge base for the reconstruction can emerge as different from an institutional rationale.

The two layers of functions and institutions can also be considered as degrees of freedom. For example, one can question whether a network at the institutional level is functionally efficient and whether it provides dynamic scale effects. (The latter can be considered as emergent synergies.) The functional perspective and the institutional perspective can be used for the optimization and the reorganization in different cycles.

## 2. The representation of a Triple Helix *dynamics*

A Triple Helix configuration can be depicted statically using social network analysis or in more general terms, as partially overlapping sets (e.g., Venn-diagrams; see also figure 2 below). While a Triple Helix dynamics can be expected to remain in flux, the geometrical representation focuses necessarily on one subdynamics or another by taking a perspective. For example, one can measure instantiations of a Triple Helix in terms of variables *or* trajectories along the time axis. An evolutionary system, however, can go through reconstructions of the complex system in the present. The system is complex at different levels, since the various subdynamics can be recombined algorithmically.

How can the values of the variables at each moment in time be related to the dynamic operation over time? Using a calculus one can study changes in the value of a variable ($x = a$) in relation to changes in the variable ($dx/dt$). Following a suggestion of Bar-Hillel (1955), Leydesdorff (1995) proposed to use information calculus for this purpose. In this study, I elaborate this calculus for Triple Helix dynamics, but let me first explain the concept of the mutual information in three dimensions using graphical representations.

Already in 1979, Goguen and Varela proposed a representation of a complex and self-organizing system using a holographic model of three interacting dynamics:

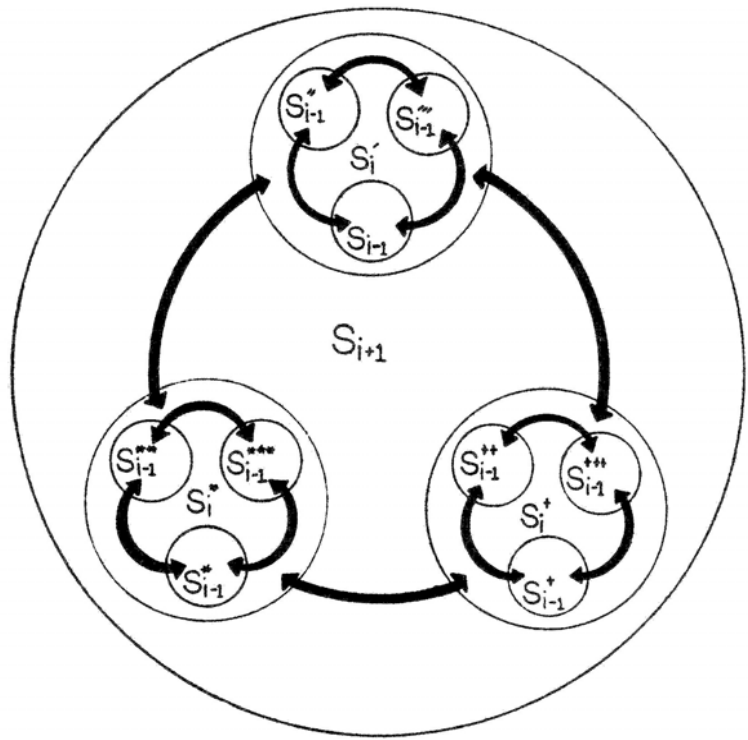

**Figure 1**
A schematic depiction of a complex system by Goguen & Varela (1979)

At each step (i-1, i, i+1), the emerging system is composed of interaction effects among the previous stages of the three participating systems. In addition, however, to the recursion of the *interaction* among the helices, a model of university-industry-government relations should encompass the *recursive dynamics within* each of the helices along their respective time axes. The differences in these subdynamics may break the symmetries suggested by this representation.

Let us develop a model that is both interactive and recursive step by step. First, consider three helices as sets that overlap in the intersections, as follows:

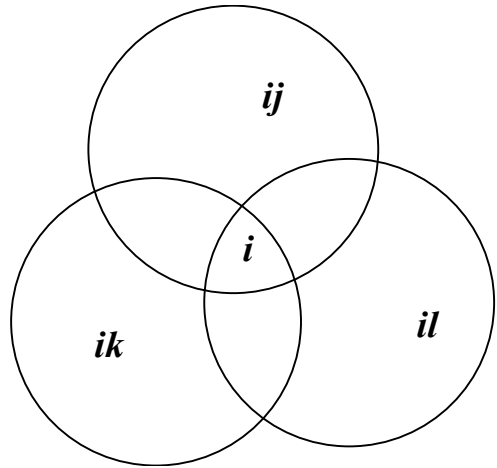

**Figure 2**
A Triple Helix configuration with positive overlap among the subsystems

In this configuration, the three helices share a common ground or origin in the overlap area indicated in the figure as *i*. Under conditions, however, this overlap can become zero or even negative. This configuration can be depicted as follows:

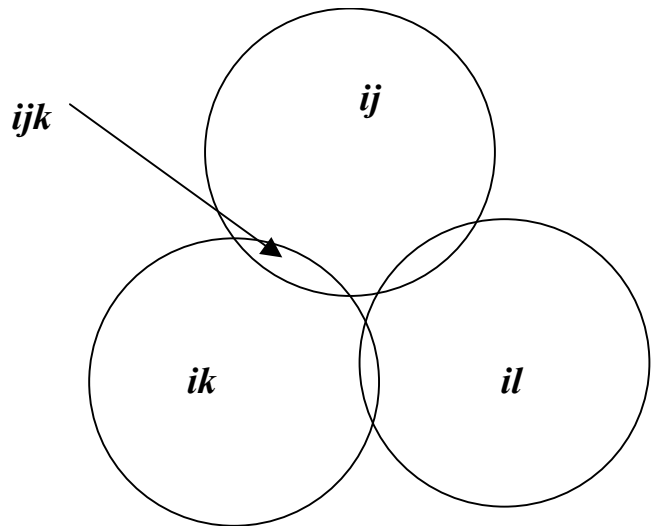

**Figure 3**
A Triple Helix configuration with negative overlap among the subsystems

In this representation, the three helices have differentiated to such an extent that the communality *i* has been dissolved. This system operates over time in terms of different communications at the respective interfaces (e.g., *ijk*). If all the interfaces operate, one can consider the result as the emergence of a 'hypercycle' (Figure 4). The hypercyclic configuration integrates the three systems in a distributed mode. It fails to integrate completely, or one can also say that the integration remains subsymbolic.

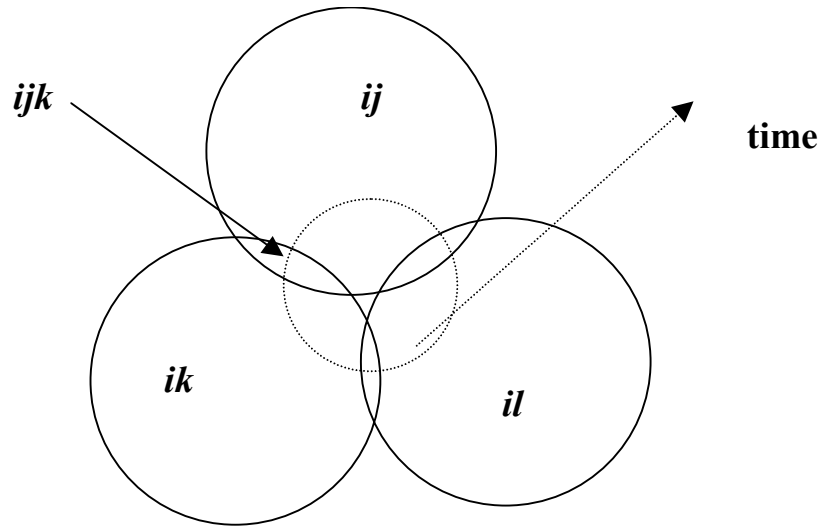

**Figure 4**
*Ex post* integration in an 'emerging' hypercycle by recombining different interactions

This configuration can be expected to exhibit 'self-organizing' properties because the various transmissions are no longer integrated at a single place. Since a common domain of instantaneous integration is lacking, each integration leads at the same time to a re-differentiation. Integration fails in this configuration at each moment in time, but it may take place over the time dimension.

It can be shown that under the condition of a lack of overlap among the three sets, the mutual information in three dimensions is negative (Abramson, 1963). From the perspective of each binary interface, the third dimension remains then 'latent' as a structural given in the background. This third system entertains interfaces with each of the first two, but not directly (or less so) with their interaction. The structural function of the third system remains beyond the control of each two relating systems, but this latent structure in the network reduces the uncertainty that prevails when the first two systems interact.

In the Triple Helix model of university-industry-government relations the hypercyclic integration can be identified as an overlay of negotiations and exchange relations among the institutional carriers of the Triple Helix dynamics. Insofar as the hypercycle operates it functions as a virtual feedback on the network of relations among the institutional agents at each moment in time.

## 4. Methodology

The mutual information in the three dimensions of the Triple Helix enables us to measure networks at each moment in time in terms of probability distributions and to evaluate the measurement results in terms of the dynamics. Unlike co-variation, correlation or co-occurrence measurements, the mutual information is defined in the case of interactions among three dimensions. However, the mutual information in three dimensions can no longer be considered as a similarity measure. It informs us about the size and the sign of the probabilistic entropy generated by the interactions within the complex system.

Conceptually, the generation of a negative entropy corresponds with the idea of complexity that is contained or 'self-organized' in a network of relations that lacks central coordination. The network system may then be able to propel itself in an evolutionary mode by alternating and recombining the various subdynamics. The reduction of the uncertainty is a result of the bi-lateral relations operating upon each other. The network contains more uncertainty-reducing structure than is visible for the interacting agents at their respective interfaces. This negative entropy is generated because the flux is constrained by the existing structure of institutional relations.

How does this relate to the measurement? Triple Helix relations can be measured in terms of relevant variables (e.g., budgets, collaborations, citations). From this perspective, the historical description of a specific configuration can be considered as measurement with only nominal variables (that is, words used for the description). In detailed ("thick") descriptions, one is able to evaluate whether something was the case or not. However, one can often specify the intensity of the relationship at a more aggregated level using measurement scales more refined than the binary one. To which extent was something the case?

For example, when comparing science parks, one may be able to count instances in which government agencies were involved in these academic-industry relations, and to which extent. In other cases, one may be able to measure more precisely, for example, along a scale. The measurement can be based on various measurement scales, but the networks can always be compared as relative frequency distributions.

Independently of the answer to the question how the network relations are operationalized and measured, the observations of Triple Helix configurations can thereafter be organized in a three-dimensional array using the format visualized in Figure 5:

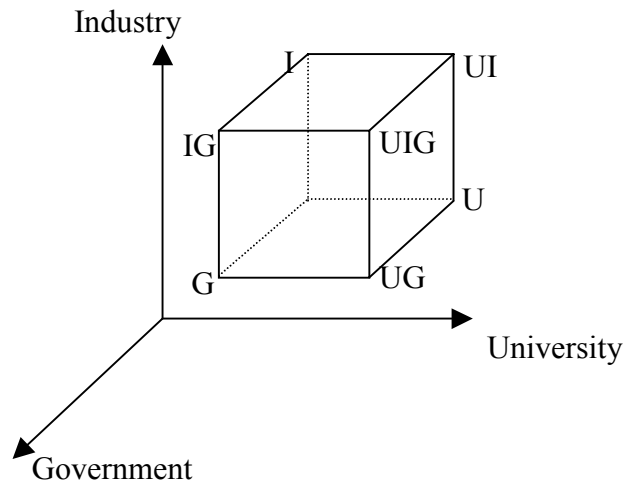

**Figure 5**
The three-dimensions of measurement in a Triple Helix configuration and their combinations

Different variables can also be measured in more than one of the three institutional dimensions. This leads to a co-variation or mutual information between the dimensions. However complicated the data gathering may be, this does not affect these methods for analyzing Triple Helix data in terms of the three dimensions indicated in Figure 5. Methodological questions about the data collections and the measurement can thus be distinguished from methodological questions with respect to the data analysis. This study focuses on the development of an indicator that can be used after and relatively independently of how the data were collected.

In general, network data can be considered as relative frequency distributions. A relative frequency distribution can be written as a probability distribution. The description of the network data in terms of probability distributions enables us to use Shannon's (1948) mathematical theory of communication. A probability distribution contains an uncertainty. The expected information content of the message that these events have happened with this observed frequency distribution, can be expressed in terms of bits of information using the Shannon-formulas (Abramson, 1963; Theil, 1972; Leydesdorff, 1995).

The mutual information between two dimensions of the probability distribution (for example, in university-industry (UI) relations) is then equal to the transmission (T) of the uncertainty (Theil, 1972):

$$T_{UI} = H_U + H_I - H_{UI}$$

The relationship reduces the uncertainty for the two relating systems (with $-H_{UI}$). Abramson (1963, at p. 129) showed that the mutual information in three dimensions can be derived as:

$$T_{UIG} = H_U + H_I + H_G - H_{UI} - H_{IG} - H_{UG} + H_{UIG}$$

Note that the uncertainty of the variables measured in each of the interacting systems ($H_U$, $H_I$, and $H_G$) is reduced at the system's level by the relations at the interfaces between them, but the three-dimensional uncertainty adds positively to the uncertainty that prevails. Because of this alteration of the signs, the three-dimensional transmission can become negative. As noted, this reduction of the uncertainty by the negative transmission is a result of the network configuration of bi-lateral relations that develops without central coordination (Figure 4).

## 5. Results

In order to show the usefulness of this indicator, I will apply it to relatively straightforward data like search results with the terms 'university,' 'industry,' 'government,' and their combinations with Boolean AND operators in various databases. As noted, the measurement problems in the data collections are backgrounded in favour of the data analysis. The data collection is based on raw search strategies that result in approximate figures, but which serve us here mainly for the illustration of the argument.

The research question behind the searches is whether and the extent to which the relations among these retrieval terms enable us to reveal a Triple Helix dynamics operating. At which level can a self-propelling dynamic of network relations be observed, and to what extent? I first turn to the Internet for retrieving relevant time-series data and then use also the *Science Citation Index* to measure these relations at national and international levels.

**5.1 The Triple Helix at the Internet**

University-industry-government relations can be measured at the Internet, for example, in terms of the occurrences and co-occurrences of the words 'university,' 'industry,' and 'government' (Leydesdorff & Curran, 2000). Using various search engines, Bar-Ilan (2001) showed, among other things, how sensitive the Internet is to measurements at different moments in time (Rousseau, 1999). However, the *AltaVista Advanced Search Engine* has remained the sole search engine that enables the analyst to combine the various search options with specific time frames (e.g., years) so that time series of data in various dimensions can conveniently be generated (Leydesdorff, 2001a).[1]

The search terms 'university,' 'industry,' 'government,' and their combinations with Boolean AND-operators were used for the years 1993-2001. All searches were performed on the 24$^{th}$ of March 2002, and during the data collection the stability of the Altavista Advanced Search Engine was checked at least once an hour for its stability (Rousseau, 1999). The results of these searches are shown in Figure 6.

---

[1] Google offers an API-service that allows for programming searches in this domain, including the Julian calendar date (Sylvan J. Katz, *personal communication*).

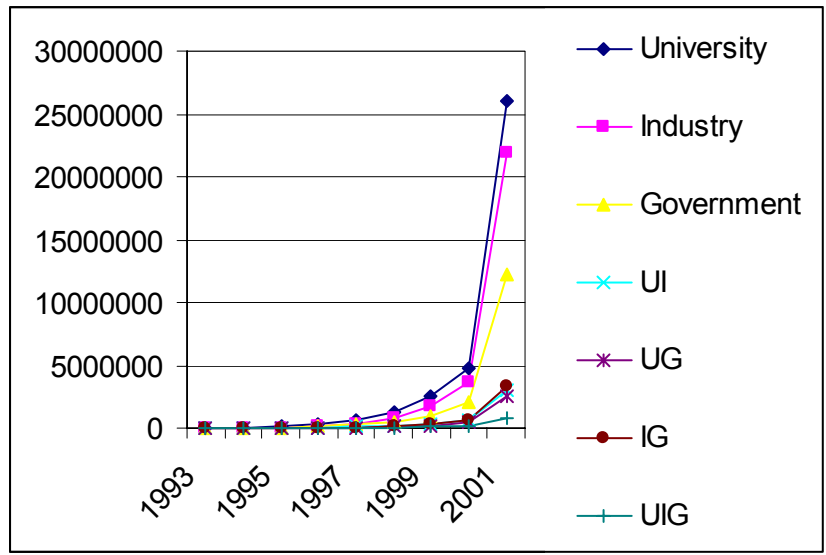

**Figure 6**
Results of searches using the *AltaVista Advanced Search Engine*

Figure 6 first shows that the Internet continues to expand rapidly. In Figure 7 the continuous growth of the number of all documents in the AltaVista domain is shown using a logarithmic scale. Remember that AltaVista provides only one specific representation of the data at the Internet (Butler, 2000; cf. Leydesdorff, 2001a).

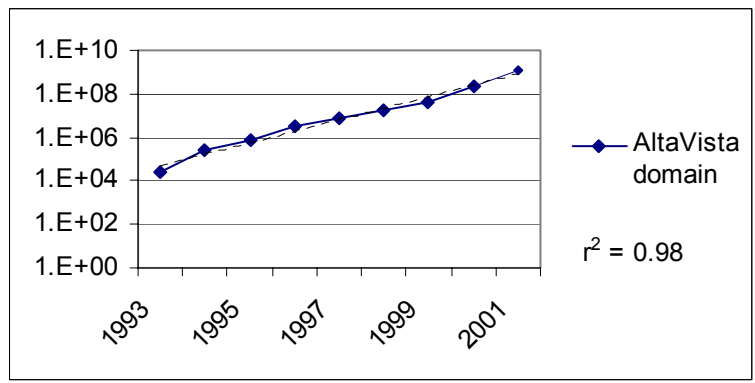

**Figure 7**
The exponential growth curve of the *AltaVista* domain during the period 1993-2001

When the data is organized in a three-dimensional array as explained above, the transmission in three dimensions T(uig) can be calculated straightforwardly for each year.[2] This leads to Figure 8.

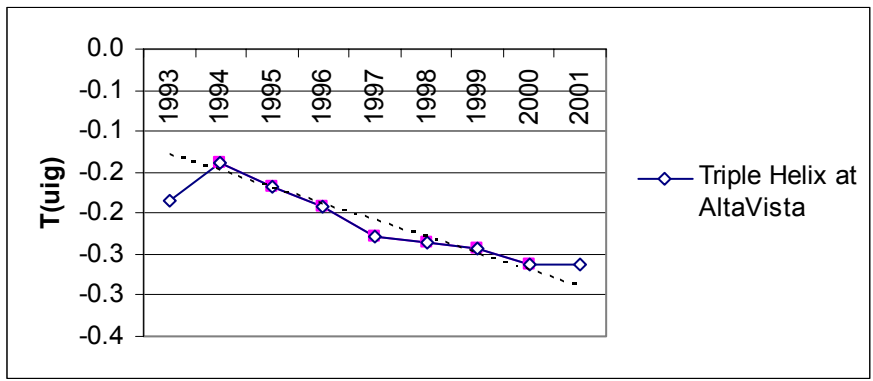

**Figure 8**
Mutual information in three dimensions ('university,' 'industry,' 'government') as measured using the *AltaVista* Advanced Search Engine. (The trend line is based on the values for 1994-2000 only; $r^2 = 0.95$.)

Figure 8 shows that the values for T(uig) are always negative, but the curve decreases linearly during the period 1994-2000. This period witnessed the booming and the potential self-organization of the so-called new economy. The decrease of the value of the transmission in three dimensions is steady during this period ($r^2 =$ 0.95). Perhaps the flattening of the curve in recent years illustrates that the process of endogenous expansion of the Internet has been interrupted temporarily as the e-business has gone into a recession. Note that this change in the dynamics is not noticeable upon visual inspection of the growth data in Figures 6 and 7.

---

[2] For the computation of the mutual information among the three dimensions, one has to assume that the search for 'university' provides us with the (margin) total for this search term. The results of the searches 'university AND industry', etc., have then to be subtracted from the total number of hits in order to find the relevant number for single occurrences of the word 'university'. The subtraction assumes Boolean consistency in the search engine, but in this case of search engines at the Internet, this condition is only statistically true. For the purposes of this study the relatively small error terms in data gathering were neglected. As noted, the focus here is on developing the indicator for the data analysis and not on improving the techniques for retrieval.

**5.2 Testing for Systemness in the Overlay of Triple Helix Relations**

What does the effect of increasingly negative values for T(uig) teach us when compared to the descriptive statistics? Does it indicate the self-organization of a virtual dimension in the overlay of relations generated by the co-occurrences? Can this, indeed, be considered as an indication of increasing self-organization of the system of relations? Are the underlying data in each of the helices also being reorganized by the emerging system at the overlay level?

Emerging systemness in data sets can be tested against the alternative of historical development of the elements of the system along the time axis (Leydesdorff, 1995). While the overlay in the Triple Helix model may exhibit systemness, the carrying institutions continue to develop historically; but the overlay system would then provide another selection environment for them at the global level, that is, in a (historically changing) present. Negative entropy first indicates that the overlay system provides the carrying systems with information relevant to reduce the uncertainty in the present. But has this feedback also become stabilized as a systemic subdynamic?

In the case of emerging systemness, one can expect a data set increasingly to contain the Markov property. The Markov property states that the current state of a system is the best prediction of its next stage.³ If systemness is not achieved, however, the normalized sum of the longitudinal predictions for the various elements provides us with the best prediction for a next state. These two hypotheses (of systemic development versus independent development of the elements, respectively) can be tested against each other for predicting next year's data. When the predicted values are subsequently observed, the quality of the two predictions can be evaluated (e.g., Leydesdorff & Oomes, 1999; Riba-Vilanova & Leydesdorff, 2001).⁴

This test was applied using the time series data 1993-2000 for the prediction of 2001 data. Comparison with the observed data for 2001 led to the following results:

---

³ Complex systems can also exhibit non Markovian features, when they have a memory. Such systems can be modeled as Markov processes with memory (Ebeling *et al*., 1995).

⁴ Since the historical prediction is dependent of the year used as the first year for the longitudinal analysis, the predictions for all possible starting year are routinely calculated and the one prediction is selected with the best fit so that the possibility of rejection of the hypothesis of systemness in the data under study is maximized.

| prediction of the value in 2001 | 7 categories (U, I, G, UI, UG, IG, UIG) | four categories (UI, UG, IG, UIG) | three categories (UI, UG, IG) |
|---|---|---|---|
| on the basis of the univariate time series (1993-2000) | 2.06 | 5.93 | 5.06 |
| on the basis of the previous year (2000) (Markov property) | 2.83 | 5.54 | 4.15 |
| hypothesis of systemness | - 0.77 (rejected) | 0.39 | 0.91 |

**Table 1**
Testing the hypothesis of systemness in the Triple Helix overlay of University-Industry-Government Relations. (All values are in millibits of information.)

The results show that the prediction of the 2001 data on the basis of the same data for the previous year (Markov property assumed) is inferior to the prediction on the basis of the time series of the various categories in the case of considering the whole system of seven search categories (second column of Table 1). Thus, the hypothesis that the representation would develop as a system is rejected.

When the analysis is limited to the three bi-lateral relations (right column of Table 1), the hypothesis of systemness in the data is corroborated. The quality of this latter prediction is worsened by including the trilateral relations (middle column). Similar results were obtained when using the prediction of data for the year 2000 on the basis of the time-series 1993-1999, but the results were then even more pronounced.[5]

---
5

In summary, these results suggest that the system of representations of university-industry-government relations at the Internet is developing as a set of bilateral relations. The bilateral relations generate a negative entropy and in this sense enable the global system to self-organize the complexity in the data using a virtual overlay of network relations. This development, however, has slowed down recently.

**5.3 The Triple Helix in the *Science Citation Index* (2000)**

In the next application of the mutual information in three dimensions on Triple Helix data, I used the 1,432,401 corporate addresses on the CD-Rom version of the *Science Citation Index 2000*. These addresses point to 725,354 records contained in this database on a total of 778,446 items. Only 3.7 % of these records contain no address information.[6] Our research focuses on the international coauthorship relations in this data, but we will report on that project elsewhere (Wagner & Leydesdorff, 2003). Here, I focus on University-Industry-Government relations in this data set.

| prediction of the value in 2000 | 7 categories (U, I, G, UI, UG, IG, UIG) | four categories (UI, UG, IG, UIG) | three categories (UI, UG, IG) |
|---|---|---|---|
| on the basis of the univariate time series (1993-1999) | 1.25 | 3.14 | 3.36 |
| on the basis of the previous year (1999) (Markov property) | 2.34 | 0.27 | 0.30 |
| hypothesis of systemness | - 0.89 (rejected) | 2.87 | 3.06 |

**Table 1a**
Testing the hypothesis of systemness in the Triple Helix overlay of University-Industry-Government Relations for the year 2000. (All values are provided in millibits of information.)

[6] The total number of authors in this database is 3,060,436. Thus, on average each record relates to four authors, but at two addresses.

An attempt was made to organize all these addresses automatically in terms of their attribution to university-industry-government relations. The routine first attributed a university label to addresses that contained the abbreviations 'UNIV' or 'COLL.' Once an attribution was made, the record was set aside before further attributions were made. The remaining addresses were subsequently labeled as 'industrial' if they contained one of the following identifiers 'CORP', 'INC', 'LTD', 'SA' or 'AG'. Thereafter, the file was scanned for the identifiers of public research institutions using 'NATL', 'NACL', 'NAZL', 'GOVT', 'MINIST', 'ACAD', 'INST', 'NIH', 'HOSP', 'HOP ', 'EUROPEAN', 'US', 'CNRS', 'CERN', 'INRA', and 'BUNDES' as identifiers.

This relatively simple procedure enabled us to identify 1,239,848, that is 86.6% of the total number of address records, in terms of their origin as 'university,' 'industry,' or 'government.' However, these results remain statistically approximate figures. The distribution is exhibited in Table 2:

|                  | *Number of records* | *Percentage* |
|------------------|--------------------:|-------------:|
| 'University'     |             878,427 |         61.3 |
| 'Industry'       |              46,952 |          3.3 |
| 'Government'     |             314,469 |         22.0 |
| – (not identified) |           192,553 |         13.4 |
| *Total*          |           1,432,401 |          100 |

**Table 2**
Number of records in the *Science Citation Index 2000* that could be attributed with a Triple Helix label using a routine

The addresses refer thus identified to 676,511 (93.3%) of the 725,354 records in the database that contain address information. Furthermore, the address information also contains the country names. For the purpose of this study, records containing an address in England, Scotland, Wales or Northern Ireland were additionally labeled 'UK,' and analogously a dataset for the EU was composed containing all records with addresses in the 15 member states. The label 'Scandinavia' was added to all records containing an address in Norway, Sweden, Denmark, and Finland. A subset of the 120,086 internationally co-authored papers could analogously be defined.

For all these subsets a three-dimensional transmission of Triple Helix relations can be calculated. The results of this calculation are shown in Table 3.

|  | number | % titles retrieved | T(uig) in mbits | UI | UG | IG | UIG | Univers | Industry | Govern |
|---|---|---|---|---|---|---|---|---|---|---|
| All | 676511 | 93.3 | **-77.0** | 16270 | 108919 | 4359 | 5201 | 543123 | 41242 | 232096 |
| USA | 232571 | 92.5 | **-74.4** | 7200 | 37834 | 1782 | 2666 | 200149 | 18154 | 66416 |
| EU | 257376 | 93.0 | **-50.1** | 4455 | 52112 | 1485 | 2028 | 206747 | 11192 | 101545 |
| UK | 68404 | 93.1 | **-63.1** | 1719 | 13098 | 394 | 690 | 54823 | 3970 | 26202 |
| Germany | 61017 | 94.7 | **-43.4** | 1028 | 14003 | 407 | 664 | 51283 | 2799 | 23701 |
| France | 41112 | 90.3 | **-52.1** | 439 | 11593 | 452 | 530 | 26133 | 1928 | 26595 |
| Scandinavia | 30939 | 95.8 | **-31.6** | 490 | 8477 | 162 | 371 | 26542 | 1263 | 13005 |
| Italy | 28958 | 89.9 | **-29.4** | 362 | 7133 | 87 | 262 | 25633 | 905 | 10526 |
| Netherlands | 18357 | 95.3 | **-25.4** | 372 | 4482 | 106 | 259 | 16379 | 863 | 6593 |
| Japan | 67715 | 97.9 | **-92.1** | 4147 | 12492 | 954 | 1311 | 56534 | 9732 | 21664 |
| PR China | 22116 | 99.5 | **-14.9** | 237 | 4610 | 68 | 114 | 18196 | 480 | 8583 |
| Taiwan | 8390 | 97.4 | **-17.1** | 148 | 2163 | 19 | 52 | 7454 | 250 | 3120 |
| Singapore | 2931 | 99.0 | **-23.9** | 104 | 476 | 7 | 17 | 2598 | 145 | 809 |
| S. Korea | 12038 | 98.3 | **-40.1** | 351 | 2341 | 87 | 91 | 10345 | 676 | 3978 |
| Russia | 22767 | 98.6 | **-24.2** | 76 | 6315 | 162 | 138 | 11507 | 478 | 17611 |
| India | 10916 | 89.2 | **-78.1** | 97 | 1813 | 61 | 55 | 6099 | 407 | 6492 |
| Brazil | 9120 | 91.0 | **-22.4** | 137 | 1727 | 32 | 52 | 7968 | 267 | 2885 |
| internationally coauthored | 120086 | 98.9 | **-21.9** | 4550 | 47054 | 1349 | 2545 | 107569 | 9422 | 61138 |

**Table 3**
University-industry-government relations for various countries and regions using ISI's *Science Citation Index 2000.*[7]

Table 3 suggests a very different pattern for the Triple Helix developments in various world regions. The Triple Helix overlay operates within the U.S.A. and Japan at a much higher level of self-organization than in Europe. Within the European Union, one can observe a scale with the U.K. at the leading end, but the smaller units (e.g., The Netherlands) at much lower levels. Russia and Brazil are even less integrated from this perspective, but India exhibits the Asian-Pacific pattern (T < –0.70).

---

[7] The numbers of records under the headings "university," "industry," and "government", respectively, in the three right-hand columns are total numbers of papers with this type of address, that is, including the papers coauthored with authors from the other sectors.

In terms of the three-dimensional transmission, Japan is by far more networked in a Triple Helix mode than the other countries included in this analysis. This can already be seen on visual inspection of the numbers. For example, the number of papers with both university and industry addresses in Japan is 4147 against 4455 for the whole EU. This corresponds to 7.3% and 2.2% of all university papers in these two subsets, respectively.

In France the ratio of university papers coauthored with industry is only 1.7%. In this case, the relation between industry and government is even stronger than the one between the university sector and industry. The relations between university research and public sector research are strong everywhere, but in France, Russia, and India public sector research is larger than university research in terms of scientific output. The East-Asian countries demonstrate how the industrial participation in the network knowledge production system can be the crucial variable for the self-organization of the Triple Helix dynamics.

Note that more than 39% of the internationally coauthored papers contain an address of both a university and a government agency. Yet, the relative low numbers of *bilateral* coauthorship relations of universities and government agencies with industrial partners indicates a low level of institutional differentiation in this subset. Furthermore there seems to be a size effect of the $T(uig)$ indicator among nations, but this correlation is not statistically significant. The main distinction, however, is visible as a pattern of collaboration that is culturally specific. The academic system on the continent in Europe seems much more traditional in its patterns of collaboration than in the U.S.A. and Japan. University-Government relations are more established in the European nations than University-Industry relations. Russia and France are the most extreme cases in this respect. The papers based on international collaborations exhibit the least development in this Triple Helix mode of relations among all the subsets.

It should be noted that these results refer to representations in the *Science Citation Index*, and the above classification into sectors was statistical and therefore approximate. Industry is weakly represented in this data. Collaborations with industry may often not lead to this type of scientific publication. The purpose of this study, however, was to demonstrate the use of three-dimensional transmissions as a methodology for data analysis. Data collection may require more care. However, independently of the refinement of the measurement, network data about university-industry-government relations can usually be written as relative frequency distributions. The indicators of the three-dimensional transmissions can then be applied to a comparison of the state of the Triple Helix configurations under study.

**5.4 Two further tests**

The above results suggest that the Triple Helix is operating strongly as an overlay in the system of global representations, but not similarly at the national level in all the advanced countries. On the contrary, cultural patterns seem to intervene. Let me therefore proceed by adding two further types of analysis, notably the representation of national systems at the level of an international database, and the representation of an international system at the level of a national database. These two tests enable me to specify the dynamics of globalization and Triple Helix relations in greater detail.

**5.4.1 National subdomains and languages at the Internet**

The Internet can be considered as a global system, but it can be searched specifically for national domains using the domain name (e.g, '.br' for Brazil) and/or the national language (Portuguese in this case). In Leydesdorff & Curran (2000) we explored these various dimensions. In this study, three national domains with their respective languages will be used: Brazil (.br) with Portuguese, Germany (.de) with German, and the Netherlands (.nl) with Dutch. Among the many possibilities, Brazil and the Netherlands were selected for the comparison with our previous analysis. Germany was added as a third case with a larger economy because Brazil and the Netherlands also exhibited some similarities in our previous analysis.

After consultation with a number of native speakers, the following search terms were selected:

|            | University    | Industry  | Government |
|------------|---------------|-----------|------------|
| *Portuguese* | Universidade | Indústria | Governo    |
| *German*     | Universit*   | Industr*  | Bundes*    |
| *Dutch*      | Universit*   | Industr*  | Overheid   |

**Table 4**
*Search terms for Triple Helix relations in different national languages*

The similarity between the English search terms and those in German and Dutch makes it possible to include the terms more globally, possibly disadvantaging Portuguese. However, the results exhibited in Figure 9 are very clear and robust: the global development at the Internet prevailed over national differences as the Internet developed during the 1990s.

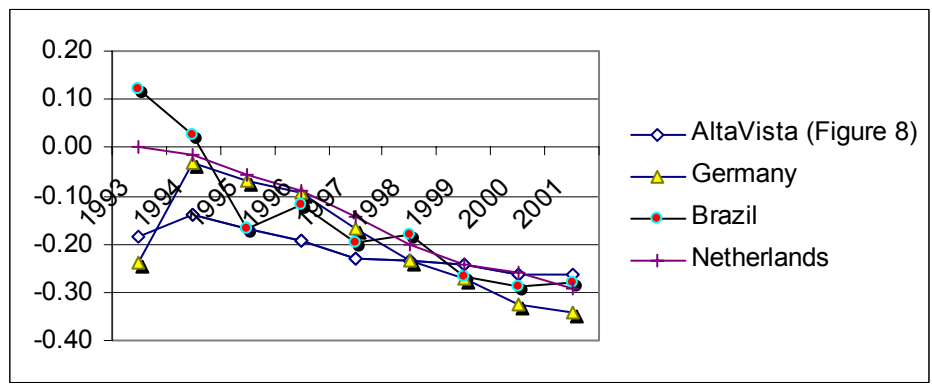

**Figure 9**
*The three-dimensional transmission between 'university,' 'industry,' and 'government' in various national systems and languages during the period 1993-2001*

Although initially (1993-1995) these 'national' representations of Triple Helix configurations were variously integrated and differentiated at the Internet, the global dynamics harmonized these systems into a similar pattern during the years thereafter. The self-organization of Triple Helix relations at the Internet has prevailed over national differences in terms of domains and languages.

**5.4.2 U.S. Patent data**

The database of the U.S. Patent and Trademark Office (USPTO) provides us with a nationally organized database which can be used as a window on international developments of patents because it integrates patent applications from around the world at the level of the American market (Narin & Olivastro, 1988; Granstrand, 1999).

'University,' 'industry,' and 'government,' and their various combinations with Boolean 'AND' operators can also be used as free text search terms in this database (Black, 2003). As in the case of the Internet, I searched the patent database for the number of occurrences of these terms in the file on a year-to-year basis.[8] For reasons of comparison with the Internet analysis in sections 5.1. and 5.4.1, the time-series in Table 5 is shown for the period 1993-2001.

---

[8] The use of the word 'industry' in a free-text search means that this word is part of the discourse, but it does not refer to the ownership or inventor of the patent in question (Debackere *et al.*, 2002). However, the number of patents assigned to corporations is much higher than indicated by 'industry' as a free-text search term.

| Year | University | Industry | Government | UI | UG | IG | UIG | Total number of patents |
|------|-----------|----------|------------|------|------|-----|------|------------------------|
| 1993 | 3063 | 9716  | 2619 | 401  | 588  | 334 | 63  | 110540 |
| 1994 | 3359 | 10568 | 2855 | 479  | 684  | 390 | 89  | 114564 |
| 1995 | 3710 | 10800 | 2828 | 529  | 771  | 410 | 93  | 114864 |
| 1996 | 4552 | 12147 | 3149 | 703  | 963  | 488 | 114 | 122953 |
| 1997 | 5406 | 12699 | 3604 | 814  | 1199 | 583 | 168 | 125884 |
| 1998 | 7623 | 17068 | 4708 | 1254 | 1658 | 807 | 266 | 166801 |
| 1999 | 8326 | 18553 | 4856 | 1352 | 1735 | 844 | 235 | 170265 |
| 2000 | 8488 | 19368 | 4831 | 1399 | 1776 | 865 | 267 | 176350 |
| 2001 | 9190 | 20812 | 5136 | 1591 | 1868 | 996 | 296 | 184172 |

**Table 5**
The number of hits for the search terms 'university,' 'industry,' and 'government' and their combinations in the database of the U.S. Patent and Trade Office.

Note that the number of patents recalled with the search term 'university' has grown steadily over the period, but this growth has declined in recent years. This breach in the trend can be made even more visible by searching for the term 'university' among the patent assignees, as is shown in Figure 10 for the whole period since the introduction of patent rights for universities by the Bayh-Dole Act of 1980.

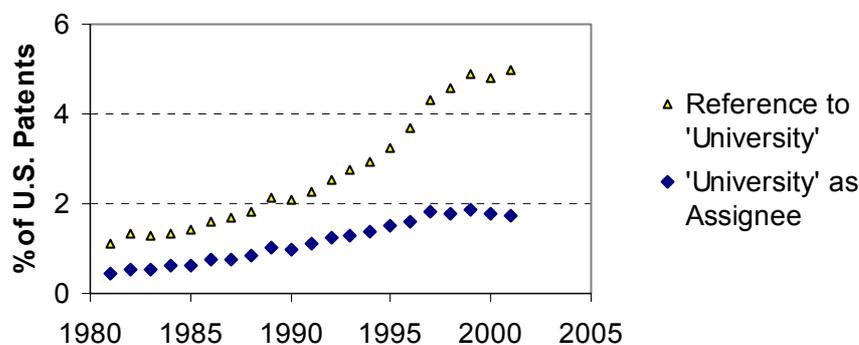

**Figure 10**
Percentage of U.S. Patents (i) with a reference to the word 'university' and (ii) a 'university' among the assignees

During the 1980s the value of three-dimensional transmissions among the dimensions 'university,' 'industry,' and 'government' has remained stable.[9] But during the 1990s this value began to rise, indicating a tendency towards a more centrally integrated and systematic word usage (Figure 11).

---
[9] During the period 1976-1992, $T_{UIG}$ had remained equal to $-0.190 \pm 0.008$.

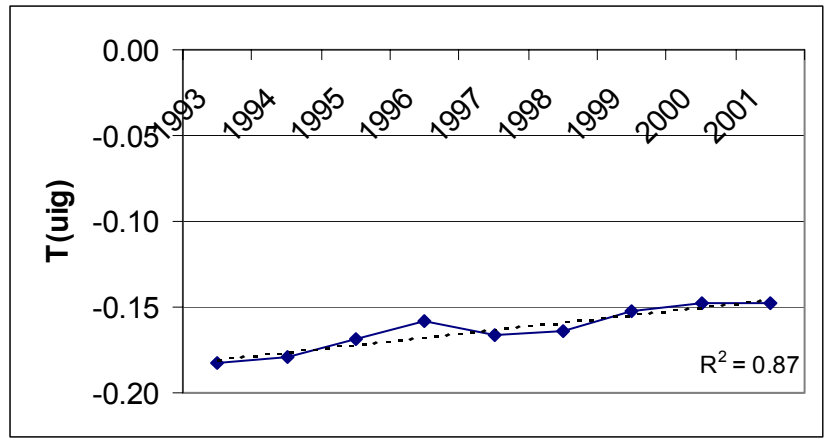

**Figure 11**
The mutual information among 'university,' 'industry,' and 'government' relations in the database of the U.S. Patent and Trade Office

With hindsight, the Bayh-Dole Act of 1980 can be considered as having provided the patent system with one more degree of freedom, that is, by allowing universities increasingly to become players in this institutional field. (The Bayh-Dole Act granted universities the right to patent results from federally funded research.) The patent system, however, has remained a highly institutionalized system of legal control and therefore under the pressure of integration. New players can be expected to be enrolled within this system in due time.

## 6. Conclusions

Triple Helix configurations can be measured as network data. The representations need by no means be restricted—like in this study—to the terms 'university,' 'industry,' and 'government' as search terms in a database. However, I have used the results from these relatively straightforward searches in order to explain how the algorithmic indicator of mutual information in three dimensions enables us to distinguish among observable Triple Helix arrangements.

For example, one can measure the discourse in an academic set of papers in terms of word (co-)occurrences and then compare the results of this analysis with a similar one for the relevant industry and government sets of papers (Leydesdorff, 2003). Matrices (or higher-order arrays) of words used in different instances, or other variables measured in the various contexts, can be compared in terms of their relative overlap. The 'mutual information' in two dimensions can be considered as a measure of the covariation. On each side, one expects remaining variations, for example, terms which are used (or have meaning) within industry, but not in academia.

If the data representation is sufficiently complex, that is, containing these three institutional dimensions, the representation can be evaluated using the three-dimensional transmission. In this study, I used scientometric and webometric representations in order to show how Triple Helix relations work differently at the national and at the global level. But the algorithm works equally well on sociometric data. The method allows us to investigate to which extent the university-industry-government networks under study exhibit the knowledge *infrastructure* of a knowledge-based *dynamics*.

The development of the dynamic knowledge base can be studied using the three-dimensional transmission because entropy can be considered as a measure of the flux. The three-dimensional transmission is generated as a potentially negative entropy within the system composed by the interacting subdynamics. Whether the overlay of the relations has also become systemic and how it then operates, for example, in terms of bilateral and/or trilateral relations can be analyzed further using information-theoretical measures (Leydesdorff, 2001b).

At the global level, the system of representations at the Internet has gone through a rapid phase of expansion and has exhibited a negative transmission in these three dimensions that is further deepening. The self-organization of the probabilistic entropy in this representation, however, was based only on the system of bi-lateral relations. The representations at the national level within this global system followed the trend and this result confirmed the assumption of a global development.

Using the ISI-database, different patterns of integration became visible in various cultural regions of the globe. Industry seems far less integrated into the academic system in Europe than in the U.S.A. and in Asia, in terms of their participation in academic publications. By using the nationally integrated U.S. patent system, I could then show that this national system exerts integrative feedback on an otherwise expanding domain. The expansion of the patent system is moderated by this institutional framework.

Finally, I indicated how the systemness of the overlay can be evaluated in comparison with the systemness of the representations of the composing units (e.g., university, industry, and government). In the case of the Internet, the bi-lateral relations between the search terms carried the systemness. The self-organization of the knowledge-based economy can thus be considered as a Triple Helix development at the global level, while nations differ in terms of how they are able to participate in these developments. The value of the mutual information in three dimensions at the national level indicates the extent to which the national system itself contains a Triple Helix dynamics.